\documentstyle[epsf]{l-aa}
\def\H{Her X-1}   \def\B{{\em BeppoSAX\/}}   \def\etal{et~al.\ }
\begin{document}

\thesaurus{08.02.03; 08.09.02 Her X-1; 08.14.1; 08.16.7 Her X-1; 13.25.3}

\title{The broad-band (0.1-200~keV) spectrum of \H\ observed with \B}

\author{D. Dal~Fiume\inst{1} \and M. Orlandini\inst{1} \and
        G. Cusumano\inst{2} \and S. Del~Sordo\inst{2} \and
        M. Feroci\inst{3} \and F. Frontera\inst{1,4} \and
        T.~Oosterbroek\inst{5} \and E. Palazzi\inst{1} \and
        A.N. Parmar\inst{5} \and A. Santangelo\inst{2} \and A. Segreto\inst{2}}

\institute{Istituto Tecnologie e Studio delle Radiazioni Extraterrestri
 (TeSRE), C.N.R., via Gobetti 101, I-40129 Bologna, Italy
\and
 Istituto Fisica Cosmica e Applicazioni all'Informatica (IFCAI), C.N.R., via
 La Malfa 153, I-90146 Palermo, Italy
\and
 Istituto di Astrofisica Spaziale (IAS),
 C.N.R., via Enrico Fermi 21, I-00044 Frascati, Italy
\and
 Dipartimento di Fisica, Universit\`a di Ferrara, via Paradiso 12,
 I-44100 Ferrara, Italy
\and
 Astrophysics Division, Space Science Department of ESA, ESTEC, Keplerlaan 1,
 2200 AG Noordwijk, The Netherlands}

\offprints{D.~Dal~Fiume - daniele@tesre.bo.cnr.it}
\date{Received [18 September 1997]; accepted [7 October 1997]}

\maketitle

\begin{abstract}
The X-ray pulsar \H\ was observed for more than two orbital cycles near the
maximum of the 35~day X-ray intensity cycle by the Narrow Field Instruments
on-board the \B\ satellite. We present the first simultaneous measurement of
the 0.1-200~keV  spectrum.  Three distinct continuum components are evident in
the phase averaged spectrum: a low energy excess, modeled as a 0.1~keV
blackbody; a power-law and a high energy cut-off. Superposed on this continuum
are Fe~L and K emission features at 1.0 and 6.5~keV, respectively, and a $\sim
40$~keV cyclotron  absorption feature. The cyclotron feature can be clearly
seen in raw count spectra. We present the properties of the cyclotron feature
with unprecedented precision and discuss the indications given by this
measurement on the physical properties of the emitting region.

\end{abstract}

\section{Introduction}

\H\ is an eclipsing binary X-ray pulsar with an  orbital period of 1.7~days
and a pulsation period of 1.2~s (Tananbaum \etal 1972; Giacconi \etal 1973). It
exhibits a 35~day cycle in X-ray intensity consisting of a $\sim 10$~day
duration main on-state and a shorter duration, a factor $\sim 3$ fainter,
secondary on-state (e.g., Gorecki \etal 1982). This modulation may result from
obscuration caused by a precessing tilted accretion disk that periodically
obscures the line of sight to the neutron star (e.g. Katz 1973; 
Petterson 1975, 1977).  In between on-states, \H\ is
still visible at a low X-ray intensity level (Jones \& Forman 1976). A regular
pattern of X-ray intensity dips are observed at certain orbital phases during
the on-states (Crosa \& Boynton 1980; Reynolds \& Parmar 1995). 

The broad band X-ray spectrum of \H\ is known to be complex. In common with
other accreting X-ray pulsars, the overall spectral shape in 1-10~keV can be
described by a power-law. In the energy range 0.1-200~keV the following
additional components have been reported: low-energy absorption consistent
with the interstellar value; a $\sim 100$~eV blackbody which dominates the
spectrum below $\sim 1$~keV (McCray \etal 1982; Oosterbroek \etal 1997); a
broad Fe~L emission feature at $\sim 1$~keV (Mihara \& Soong 1994;
Oosterbroek \etal 1997); a broad Fe~K
emission feature at $\sim 6.5$~keV  which was studied in detail by Choi \etal
(1994) using {\em Ginga\/} data; an approximately exponential cut-off to
the  power-law $>10$~keV  and a broad feature, centered around 40~keV,
which is usually interpreted as a cyclotron absorption or emission feature
(Tr\"umper \etal 1978; Voges \etal 1982; Soong \etal 1990; Mihara \etal 1990).
\H\ was the first cosmic X-ray source from which a cyclotron feature was
detected (Tr\"umper \etal 1978).

In this {\it Letter\/} we present the first simultaneous measurement  of the
\H\ spectrum over the 0.1-200~keV energy range. The Narrow Field
Instruments (NFI) on-board \B\ (Boella \etal 1997a) consist of the imaging
Low-Energy Concentrator Spectrometer (LECS; 0.1-10~keV; Parmar \etal 1997),
the imaging Medium-Energy  Concentrator Spectrometer (MECS; 1.5-10~keV; Boella
\etal 1997b),  the High Pressure Gas Scintillation Proportional Counter
(HPGSPC;  3-150~keV; Manzo \etal 1997) and the Phoswich Detector System (PDS;
15-300~keV; Frontera \etal 1997).

\section{Observation}

\B\ observed \H\ between 1996 July 24 00:34 UT and  July 27 11:54 UT. Source
spectra were extracted excluding intervals when \H\ was in eclipse or
undergoing X-ray intensity dips. The total source exposure times are 36~ks for
the LECS, 90~ks for the  MECS, 26~ks for the HPGSPC, and 47~ks for the PDS. The
different exposure times are due to rocking collimators (HPGSPC and
PDS), to operative time only during satellite night time (LECS) and
to different filtering criteria during passages in the South Atlantic 
Geomagnetic Anomaly and before and after Earth occultations.

Standard data extraction techniques were used for all instruments. That used
for the LECS is described in Oosterbroek \etal (1997). In LECS and MECS
\H\ is a factor $>100$ times the background counting rate and so
systematic effects in the background subtraction are negligible.
The methods of background subtraction
using the rocking collimator technique  for the HPGSPC and PDS are described in
Frontera \etal (1997) and Manzo \etal(1997). The standard collimator rocking
angles of 210$'$ for the PDS and 180$'$ for the HPGSPC were used together with
a dwell time for each on- and off-source position of 96~s. The systematic error
in PDS background subtraction is negligible for sources whose flux is higher
than 1~mCrab, including \H\ (Guainazzi \& Matteuzzi 1997).

\subsection{The X-ray spectrum}

The phase averaged X-ray spectrum was modeled using seven previously observed
components: (1) a low-energy excess described by a blackbody of temperature
kT${\rm _{BB}}$, (2) a Gaussian feature  with energy ${\rm E_{L}}$~keV, and
full width half-maximum (FWHM) of ${\rm FWHM_{L}}$~keV, (3) low-energy
absorption, ${\rm N_H}$, using the coefficients of Morrison \& McCammon (1983)
together with  the elemental abundances of Anders \& Ebihara (1982), (4) a
broken power-law continuum with photon indices $\alpha_1$ and $\alpha_2$ and
break energy  E$_{\rm break}$, (5) an Fe~K emission feature, (6) a high energy
cut-off to the broken power-law and (7) a cyclotron feature. Among these
components, the shapes of the high energy cut-off and cyclotron feature are
most uncertain. Version 9.0 of
the XSPEC spectral fitting program (Arnaud 1996) was used. A total of 15-19
parameters (depending on how the above components were modeled) were allowed to
vary during the fits. In addition, the normalization of each of the NFI was
allowed to vary in order to help account for the known calibration 
uncertainties between the instruments. In Table~\ref{tab:fits} we report the
result of the spectral fits using the above seven components and a 
multiplicative Gaussian (Model 1 - Soong \etal 1990) or a Lorenzian 
(Model 2 - Mihara \etal 1990) absorption line to
model component 7. Unsatisfactory fits were obtained using an emission line.

\begin{figure*}
\vspace{6.5cm}
\centerline{\includegraphics{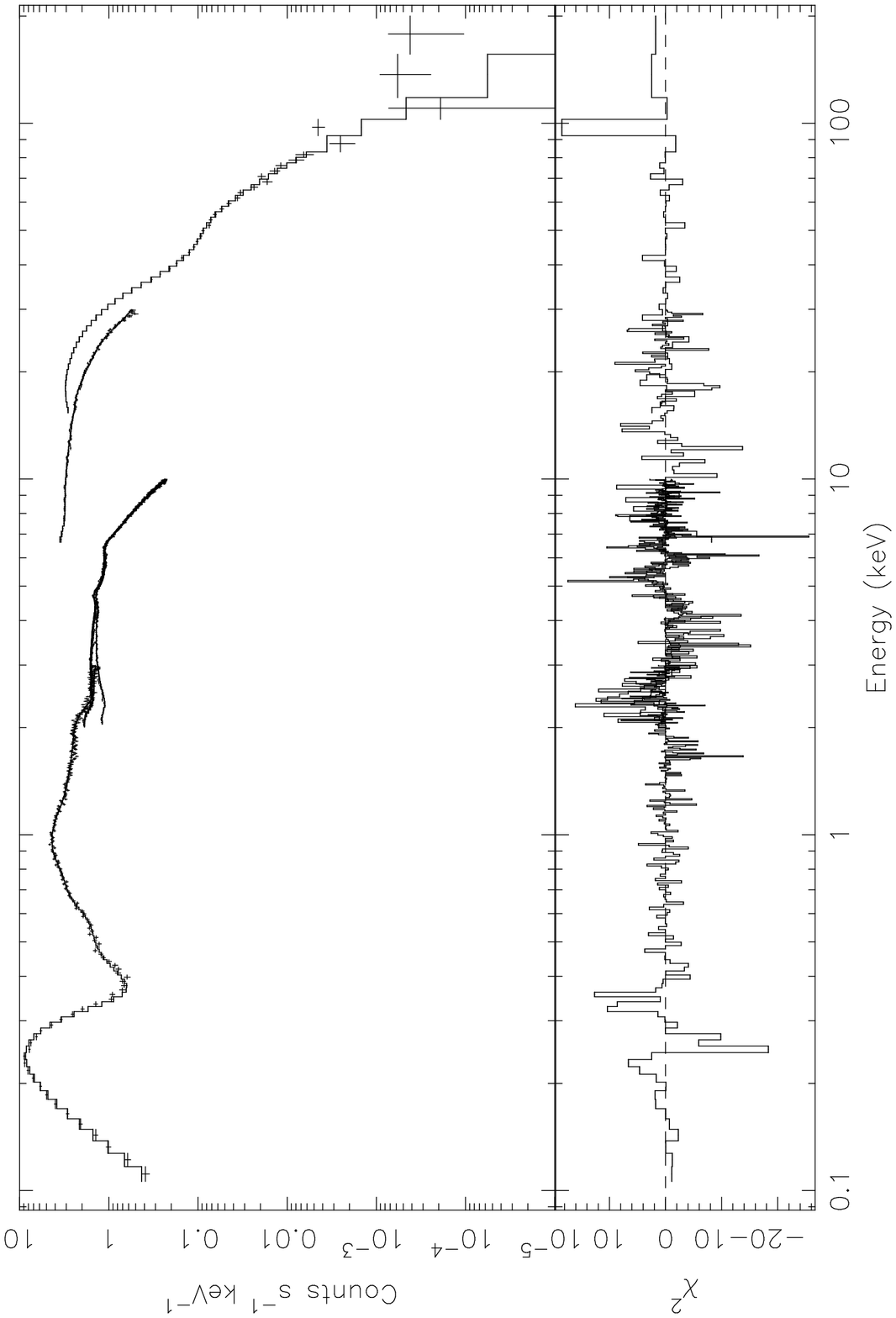}
\includegraphics{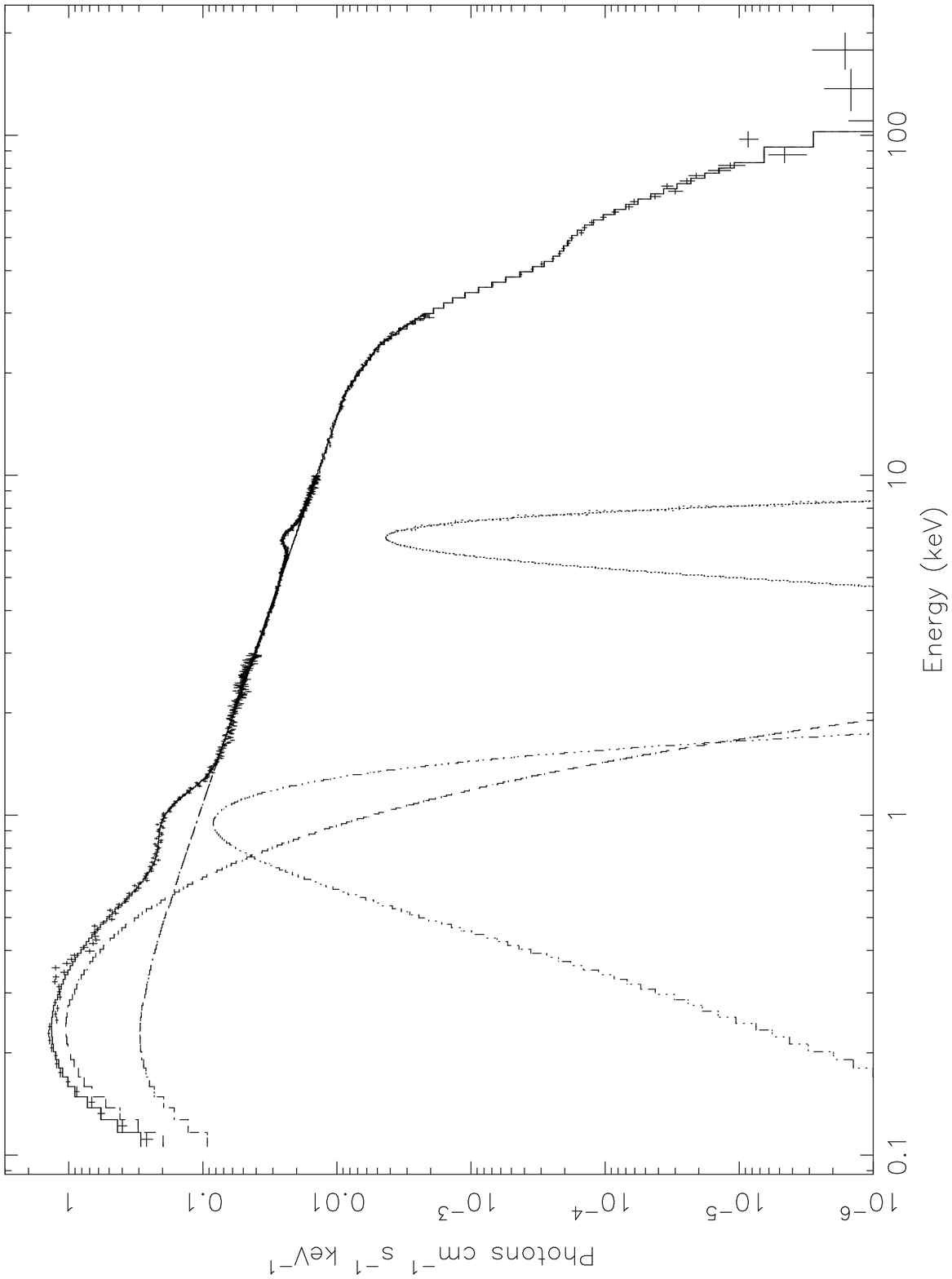}}
\caption[]{The spectrum of \H\ in the 0.1-200~keV energy  range observed by
the \B\ NFI. Left panel: count rate spectrum and contribution to $\chi^2$. The
best-fit obtained using model 1 is shown as a histogram. Right panel:
deconvoluted photon spectrum. The different spectral components used in the fit
are indicated by the dashed lines.}
\label{fig:spectra}
\end{figure*}

\begin{table}
\caption{Spectral fit parameters}
\begin{flushleft}
\begin{tabular}{lll}
\hline\noalign{\smallskip}
 Parameter & Model 1 & Model 2 \\
\noalign{\smallskip}
\hline\noalign{\smallskip}
 ${\rm I_{pow}} (\times 10^{-1}$~photons~cm$^{-2}$~s$^{-1}$)
                                             & 1.083$\pm$0.006 & 1.081$\pm$0.006 \\
 $\alpha _1$                                 & 0.884$\pm$00.003 & 0.882$\pm$0.003 \\
 $\alpha _2$                                 & 1.83$\pm_{0.07}^{0.05}$ & 1.77$\pm$0.07 \\
 ${\rm E_{break}}$ (keV)                     & 17.74$\pm^{0.26}_{0.30}$ & 17.86$\pm^{0.21}_{0.32}$ \\
 ${\rm I_{BB}}$ (L$_{36}$/d$_{5}^2$)         & 0.66$\pm 0.02$ & 0.66$\pm 0.02$ \\
 ${\rm kT_{BB}}$ (keV)                       & 0.092$\pm$0.002 & 0.092$\pm$0.002 \\
 ${\rm N_H}(\times 10^{19}$~atoms~cm$^{-2}$) & 5.1$\pm$0.7 & 5.1$\pm$0.7 \\ 
 ${\rm I_{L}}(\times 10^{-2}$~photons~cm$^{-2}$~s$^{-1}$)
                                             & 3.53$\pm{0.24}$ & 3.54$\pm{0.24}$ \\
 ${\rm E_{L}}$ (keV)                         & 0.945$\pm$0.012 & 0.945$\pm$0.012 \\
 ${\rm FWHM_{L}}$(keV)                       & 0.390$\pm$0.024 & 0.390$\pm$0.024 \\
 ${\rm I_{K}}(\times 10^{-3}$~photons~cm$^{-2}$~s$^{-1}$)
                                             & 4.88$\pm 0.18$ & 4.84$\pm_{0.16}^{0.19}$ \\
 ${\rm E_{K}}$ (keV)                         & 6.55$\pm$0.015 & 6.55$\pm$0.015 \\
 ${\rm FWHM_{K}}$ (keV)                      & 1.06$\pm$0.06 & 1.06$\pm$0.06 \\
 ${\rm E_{cut}}$ (keV)                       & 24.2$\pm$0.2 & 24.3$\pm$0.3 \\
 ${\rm E_{fold}}$ (keV)                      & 14.8$\pm$0.4 & 15.2$\pm$0.4 \\
 ${\rm E_{cycl}}$ (keV)                      & 42.1$\pm$0.3 & 40.3$\pm$0.2 \\
 ${\rm EW_{cycl}}$(keV)                      & 14.9$\pm_{1.0}^{1.25}$ & - \\
 ${\rm FWHM_{cycl}}$ (keV)                   & 14.7$\pm$0.9 & 16.3$\pm$1.4 \\
 ${\rm Depth_{cycl}}$                        & - & 0.73$\pm$0.03 \\
 $\chi^2_\nu$ (dof)                          & 1.984 (908) & 1.987 (908) \\
\noalign{\smallskip} \hline \noalign{\smallskip}
\multicolumn{3}{l}{{\sc Note} -  Errors correspond to 90\% single parameter
confidence level} \\
\end{tabular}
\label{tab:fits}
\end{flushleft}
\end{table}

Given the number of spectral components observed from \H\ in different energy
ranges by different missions, we expected that a single fit to the the entire
0.1-200~keV spectrum would be problematic. For the high energy cut-off and
cyclotron feature we tried many of the currently used empirical or
semi-empirical models, but none of them gave acceptable values of
$\chi^2_\nu$. We can exclude the single harmonic cyclotron absorption model of
Mihara \etal (1990) without high energy exponential cutoff and the Fermi-Dirac
Cut-Off (FDCO) plus cyclotron absorption model (Mihara 1995), as they gave very
high values of $\chi^2_\nu$.  We also obtained worse results using a single
(not broken) power law. We report in Table~\ref{tab:fits} only the results from
the best two model fits. These are (1) a broken power law plus high energy
exponential cut-off together with a Gaussian absorption feature and (2) a
broken power-law plus a high energy exponential cut-off together with a
Lorenzian absorption feature (Mihara \etal 1990). The two models give
comparable results in term of $\chi^2_\nu$. While the fit results are still not
formally acceptable, given the known systematic uncertainties in the NFI of
$\sim 5$\%, we do not think it worthwhile adding extra components.
The total observed flux at earth in the 0.1-200 keV energy band is 7.2$\times
10^{-9}$~erg~cm$^2$~s$^{-1}$ that corresponds to a total observed luminosity of
\H\ $\rm L_x=2.15\times 10^{37}$ erg~s$^{-1}$, assuming a distance of 5~kpc.

\begin{figure}
\vspace{8.5cm}
\centerline{\includegraphics{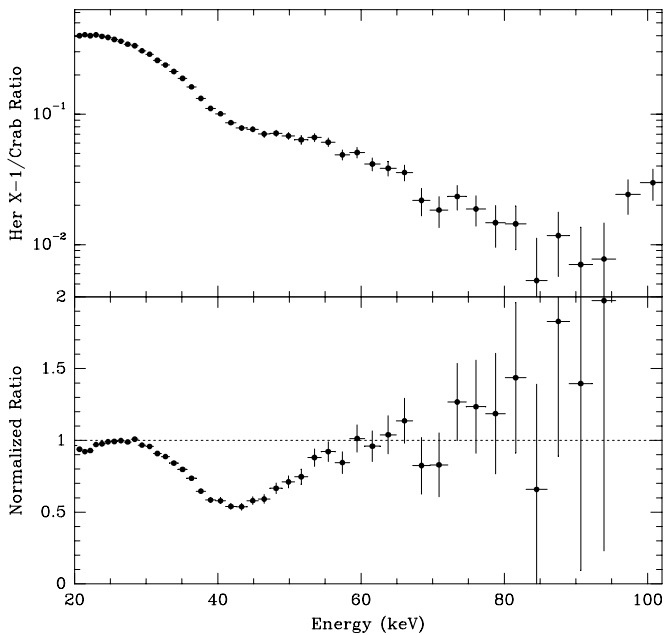}}
\caption[]{Upper panel: ratio of the 20-100~keV PDS count rate spectra of \H\
and the Crab Nebula. Lower panel: the same ratio normalized to the \H\
continuum (see text).}
\label{fig:crabratio}
\end{figure}

The broad band spectrum of \H\ appears to be even more complex  than expected.
In particular, the high energy cutoff cannot be modeled  satisfactorily using a
single power law, neither adding the ``classical'' exponential cutoff (White
\etal 1983), nor adding the more recent cyclotron absorption model of Mihara
\etal (1990). This can be clearly seen in the 10-20~keV residuals  where large
structured variations are present using these models. The use of a broken power
law, with an exponential cutoff near 24 keV greatly improves the fit as can be
seen from the fit residuals shown in Fig.~\ref{fig:spectra}.

%
The cyclotron feature is clearly visible in the raw count rate spectra. In
Fig.~\ref{fig:crabratio} (upper panel), the ratio between the \H\ and the Crab
Nebula count spectra is shown (the Crab Nebula spectrum is expected to be
featureless in this energy band). The overall shape of this ratio is dominated
by the difference in the shape of the two continua, \H\ being harder below
$\sim 23$~keV. A clear deviation from the smooth ratio of the two continua is
seen as a broad hump that extends from 40 to 100~keV. This does not have the
shape of a narrow emission or absorption line, as expected from previous
measurements. Given that the ratio of the spectra is largely independent of the
uncertainties in the energy calibrations and spectral reconstruction, it can be
also used to derive a model and calibration-free estimate of the energy at
which the \H\ spectral shape begins to deviate from the continuum.

In order to more clearly see the shape of the high energy absorption feature,
the count rate ratios were multiplied by the slope of the Crab Nebula photon
spectrum ($E^{-2.1}$). Then we divided by the
continuum functional, namely the broken power law plus high-energy cutoff,
which approximates the \H\ high energy continuum {\it without\/} the cyclotron
feature. The effect of these two steps is to obtain a normalized spectrum that
approximates the actual photon spectrum, as the division by the Crab Nebula
spectrum removes, to first order, the instrumental effects. The effect of the
second step is to enhance the deviations of this normalized spectrum from the
continuum shape described by the best-fit model. In the continuum functional we
used the parameters obtained from the XSPEC fit, model 1 (see
Table~\ref{tab:fits}). We obtained an identical result using the continuum
functional of model 2.
The lower panel of Fig.~\ref{fig:crabratio} shows a broad, symmetric absorption
line, centered just above 40 keV. As expected, this is in good agreement with
the fit parameters listed in Table~\ref{tab:fits}. The count ratio returns
consistent with unity above $\sim 55$~keV. For a cyclotron line energy of
42~keV the magnetic field strength is 3.5$\times 10^{12}$~G.

\section{Discussion}

The low-energy (E$<2$~keV) part of the spectrum is discussed in Oosterbroek
\etal (1997) and will not be discussed further here. The Fe K line measurement
confirms the findings of Choi \etal (1994). We measure a similar line energy
and width. Our spectral resolution, while a factor $\sim 3$ better than that of
{\em Ginga\/} LAC, does not allow the structure of the line to
be further investigated. A slightly asymmetric shape supports the hypothesis of
Choi \etal that the feature is a blend originating from multiple lines at
different degrees of ionization.

The underlying shape of the continuum of X-ray pulsars is still an open
question. Following the first studies of scattering and opacities in strongly
magnetized plasmas (e.g. Canuto \etal 1971)
various attempts to predict the shape of the emitted X-ray
spectrum have been made. Monte-Carlo methods were used by M\'esz\'aros \& Nagel
(1985a, 1985b, hereafter MN-I and MN-II) to calculate spectra, cyclotron line 
shapes and pulse profiles. They use a magnetic field intensity of 3.3$\times
10^{12}$~G, which gives a cyclotron energy of 38~keV. This choice is close to
our measured value in the case of \H, and therefore their results can be
readily compared to our spectrum.  Our measurement strongly indicates that the
cyclotron feature is in absorption, in agreement with the model calculations. 
The model predicts a line FWHM ($\Delta\omega_B$), dependent on the angle
between the line of sight and the field direction, $\theta$, broader for
smaller angles. Using Eq.~13 in MN-I,
\begin{equation}
\Delta \omega_B \simeq \omega_B \left( 8 \times \ln(2) \times 
\frac{\rm kT_e}{\rm m_ec^2} \right)^{\frac{1}{2}} |\cos\theta|
\label{eq13}
\end{equation}
where $\omega_B$ is the cyclotron line frequency, and $\rm m_ec^2$ is the
electron rest mass, and assuming $|\cos\theta|\approx 1$, we obtain a lower
limit of the  electron temperature ${\rm kT_e}$ of approximately 11~keV (we
used a line width of $\sim 15$~keV - see Table~\ref{tab:fits}). This is in
fair agreement with the calculations of self-emitting atmospheres of Harding
\etal (1984). In the case of a magnetic field of 3.5$\times 10^{12}$~G the
predicted  temperature for a slab or column with optical depth of the order of
50 g cm$^{-2}$ (M\'esz\'aros \etal 1983) is approximately 4-8$\times 10^7\
^\circ$K, which corresponds to ${\rm kT}_e \sim 3.5$-7~keV.
Harding \etal (1984) predict higher temperatures for higher  magnetic fields.
This is also in agreement with \B\ observations of \H\ and Vela X-1 (Orlandini
\etal 1997). Indeed, in the case of Vela X-1 the cyclotron energy is 60~keV
and the FWHM is $\sim 26$~keV, corresponding to an electron temperature of
approximately 18~keV. The harder spectrum observed from Vela X-1 also confirms
the correlation between spectral hardness and magnetic field intensity (see
also Frontera \& Dal Fiume 1989; Makishima \etal 1990). The low energy
power-law is reproduced in MN-I and MN-II qualitatively, and has a spectral
photon index somewhat flatter than our observed value of $\sim -0.8$.

More recent models on line profile and widths (Araya and Harding, 1996) were
obtained for A0535+26, using a magnetic field intensity much higher than that
measured in \H. No analytical model neither for the spectrum nor for the line
profile is available, therefore their results cannot be compared with our
measurement. They caution that the ``classical'' estimate of electron
temperature using line width, as in Eq.~\ref{eq13}, can be flawed and suggest
an estimate taken from Lamb \etal (1990)
\begin{equation}
\rm T_e = \frac{\omega_B}{\left(2+\alpha_{eff}\right)}
\label{eqHarding}
\end{equation}
that is valid in the limit of a single scattering. In this relation
$\alpha_{\rm eff}$ is the effective power law slope at the cyclotron line. The
spectrum of \H\ cannot be approximated above 20 keV with a power-law, but a
rough estimate of the slope gives values of 3.5-5. Assuming the centroid of
the cyclotron feature at 42~keV we obtain, from Eq.~\ref{eqHarding}, an
estimate of the electron temperature T$_{\rm e}$ of 6-7.6~keV.

The empirical models used here to fit the broad band spectrum of \H\ are
clearly inadequate to extract all the information about the physical properties
of the emitting region. However this work illustrates the importance of broad
band studies of X-ray pulsars in gaining insight into the physical processes
of the emission region. In particular, better analytical models of both
continuum and lines can give better and more reliable estimates of the physical
properties of the emitting region as electron temperature, density and size of
the emitting region.
\begin{acknowledgements} We thank the \B\ Science Team, the \B\ SDC and
the Nuova Telespazio staff at the \B\ Control Center for their support. 
This research was in part supported by a
dedicated grant of Agenzia Spaziale Italiana.
\end{acknowledgements}


\begin{thebibliography}{}

\bibitem[1982]{abundance}
Anders, E., Ebihara, M. 1982, Geochim. Cosmochim. Acta, 46, 2363

\bibitem[1996]{araya2}
Araya, R.A., Harding, A.K. 1996, ApJ, 463, L33

\bibitem[1997]{xspec}
Arnaud, K.A., 1996, In: Jacoby G., Barnes J. (eds.) Astronomical Data
Analysis Software Systems V, ASP Conf. Series, 101, p.17

\bibitem[1997]{sax}
Boella, G., Butler, R.C., et~al. 1997a, A\&AS, 122, 299

\bibitem[1997]{mecs}
Boella, G., Chiappetti, L., et~al. 1997b, A\&AS, 122, 327

\bibitem[1971]{canuto}
Canuto, V., Lodenquai, J., Ruderman, M. 1971 Phys. Rev., D3, 2303

\bibitem[1994]{choi}
Choi, C.S., Nagase, F., Makino, F. 1994, ApJ, 437, 449

\bibitem[1980]{crosaboynton}
Crosa, L., Boynton, P.E. 1980, ApJ 235, 999

\bibitem[1997]{pds}
Frontera, F., Costa E., et~al. 1997, A\&AS, 122, 357

\bibitem[1989]{ffddf}
Frontera, F., Dal~Fiume, D. 1989, In: Proc. 23$^{\rm th}$ ESLAB Symposium, ESA
 SP-296, p.37

\bibitem[1973]{giacconi}
Giacconi, R., Gursky, H., et~al. 1973, ApJ, 184, 227

\bibitem[1982]{gorecki}
Gorecki, A., Levine, A., et~al. 1982, ApJ, 56, 234 

\bibitem[1997]{pdsbkg}
Guainazzi, M., Matteuzzi, A. 1997, BSAX-SDC TR-011

\bibitem[1984]{harding84}
Harding, A.K., M\'esz\'aros, P., et~al. 1984, ApJ, 278, 369

\bibitem[1990]{jones}
Jones, C., Forman, W. 1976, ApJ, 209, L131

\bibitem[1973]{katz}
Katz, J.I. 1973, Nature Phys. Sci., 246, 87

\bibitem[1990]{lamb}
Lamb, D.Q., Wang, J.C.L., Wasserman, I.M. 1990, ApJ, 363, 670

\bibitem[1990]{makish}
Makishima, K., Mihara, T., et~al. 1990, ApJ, 365, L59.

\bibitem[1997]{hp}
Manzo, G., Giarrusso, S., et~al. 1997, A\&AS, 122, 341

\bibitem[1982]{mccray}
McCray, R.A., Shull, J.M., et~al. 1982, ApJ, 262, 301

\bibitem[1983]{meshard}
M\'esz\'aros, P., Harding, A.K., et~al. 1983, ApJ, 266, L33

\bibitem[1985]{meszaros3}
M\'esz\'aros, P., Nagel, W. 1985a, ApJ, 298, 147 (MN-I)

\bibitem[1985]{meszaros4}
M\'esz\'aros, P., Nagel, W. 1985b, ApJ, 299, 138 (MN-II)

\bibitem[1990]{miharath}
Mihara T. 1995, Ph.D. Thesis, Riken.

\bibitem[1990]{mihara}
Mihara, T., Makishima, K., et~al. 1990, Nat, 346, 250

\bibitem[1983]{morrison}
Morrison, R., McCammon, D. 1983, ApJ, 270, 119

\bibitem[1997]{velax1}
Orlandini, M., Dal~Fiume, D., et~al. 1997, A\&A, submitted

\bibitem[1997]{ooser97}
Oosterbroek, T., Parmar, A.N., et~al. 1997, A\&A, in press

\bibitem[1997]{lecs}
Parmar, A.N., Martin, D.D.E., et~al. 1997, A\&AS, 122, 309

\bibitem[1975]{petterson1}
Petterson, J.A., 1975 ApJ, 201, L61

\bibitem[1975]{petterson2}
Petterson, J.A. 1977, ApJ, 218, 783

\bibitem[1995]{reypar}
Reynolds, A.P., Parmar, A.N. 1995, A\&A, 297, 747

\bibitem[1990]{soongcy}
Soong, Y., Gruber, D.E., et~al. 1990, ApJ, 348, 641

\bibitem[1972]{tananbaum}
Tananbaum, H., Gursky, H., et~al. 1972, ApJ, 174, L143

\bibitem[1978]{trumpcy}
Tr\"umper, J., Pietsch, W., et~al. 1978, ApJ, 219, L105

\bibitem[1982]{voges}
Voges, W., Pietsch, W., et~al. 1982, ApJ, 263, 803

\bibitem[1983]{whiswah}
White, N.E., Swank, J.H., Holt, S.S. 1983, ApJ, 270, 711

\end{thebibliography}
\end{document}